\documentclass[twocolumn,aps,prb,amsmath,amssymb,superscriptaddress]{revtex4-1}    
\usepackage{amsfonts}
\usepackage{amssymb}
\usepackage{graphicx}
\usepackage{dcolumn}
\usepackage{bm}
\usepackage{ulem}
\usepackage{color}

\begin{document}

\title{\textit{Ab initio} study of topological surface states of strained HgTe}

\author{Shu-Chun Wu}
\affiliation{Max Planck Institute for Chemical Physics of Solids, D-01187 Dresden, Germany}
\author{Binghai Yan}
\email{yan@cpfs.mpg.de}
\affiliation{Max Planck Institute for Chemical Physics of Solids, D-01187 Dresden, Germany}
\affiliation{Max Planck Institute for Physics of Complex Systems, D-01187 Dresden, Germany}
\author{Claudia Felser}
\affiliation{Max Planck Institute for Chemical Physics of Solids, D-01187 Dresden, Germany}

\date{\today}

\begin{abstract}
The topological surface states of mercury telluride (HgTe) are studied by \textit{ab initio} calculations
assuming different strains and surface terminations. For
the Te-terminated surface, a single Dirac cone exists at the $\Gamma$ point. The Dirac
point shifts up from the bulk valence bands into the energy gap when the substrate-induced
strain increases. At the experimental strain value (0.3\%), the Dirac
point lies slightly below the bulk valence band maximum. A left-handed spin texture was
observed in the upper Dirac cone, similar to that of the Bi$_2$Se$_3$-type topological
insulator. For the Hg-terminated surface, three Dirac cones appear at three
time-reversal-invariant momenta, excluding the $\Gamma$ point, with nontrivial spin textures.

\end{abstract}
\maketitle


\section{introduction}

Topological insulators (TIs) are a new quantum state of matter that is attracting much
attention in condensed matter physics for both its fundamental interest and
its potential applications.~\cite{qi2010,moore2010,hasan2010,qi2011RMP} Mercury telluride
(HgTe) is a prototype of a TI material with an inverted band
structure~\cite{bernevig2006d,koenig2007} that distinguishes it from a normal insulator.
However, its conduction and valence bands are degenerate at the Fermi surface because of
the cubic symmetry, inducing a zero energy gap. Breaking the cubic symmetry can
open a band gap and realize a true TI in two ways. One is by reducing the dimensions from three
to two in thick CdTe/HgTe/CdTe quantum wells (QWs).~\cite{bernevig2006d} This is the
first experimentally realized TI system,~\cite{koenig2007} which is also called the
quantum spin Hall effect. The other way is by using an external
strain~\cite{fu2007a,dai2008,bruene2011} (e.g., from the CdTe substrate) to induce a three-dimensional (3D) TI.
Two-dimensional (2D) HgTe QWs have been extensively studied as a model system for phenomena including edge state
conductance,~\cite{roth2009} spin polarization,~\cite{Bruene2012} and even direct
imaging.~\cite{Ma2012} In contrast, 3D HgTe has not been explored as fully
as other 3D TIs such as Bi$_2$Se$_3$.~\cite{Zhang2009,xia2009} Although quantum Hall
measurements and angle-resolved photoemission spectroscopy (ARPES)~\cite{bruene2011}
recently verified the existence of topological surface
states (TSSs), the surface state dispersion above the Dirac point and the
helical spin texture have yet to be explored. This is plausible for two reasons: (i)
in experiments, strained HgTe samples, which are not easily fabricated, are usually hole-doped,
which hinders the access of ARPES to unoccupied surface states; (ii) in theory, the
requirement of thick slab models significantly increases the computational cost of
\textit{ab initio} calculations.

In this article, we studied the TSSs of strained HgTe as a 3D TI and focused on their
spin textures by using maximally localized Wannier functions (MLWFs)~\cite{Mostofi2008}
extracted from \textit{ab initio} calculations, which enables us to simulate large surface
models. Although the existence of TSSs is protected by the topology of the bulk bands, their
energy dispersion is found to depend sensitively on the external strain and local atomic
structures of the surface. The helical spin textures of the Dirac-type surface states
are revealed, where the upper Dirac cone exhibits a left-hand chirality with a Berry
phase $\pi$.


\section{methods}

\begin{figure}[htb]
    \includegraphics[width=1\linewidth]{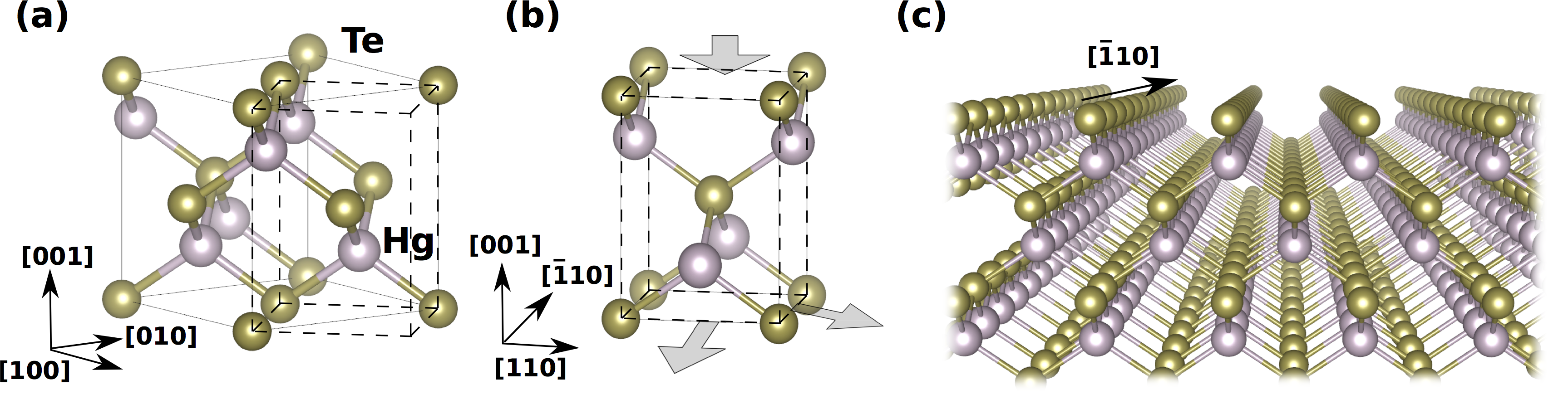}
    \caption{(a) Zinc blende crystal structure of HgTe in a cubic unit cell. White and green balls represent Hg and Te atoms, respectively. Dashed lines indicate a tetragonal unit cell. (b) Tetragonal unit cell. Substrate-induced strains are applied as in-plane tensile strains and a uniaxial compressive strain. (c) Te atoms on the top of the Te-terminated surface form a 1D chain along the [\={1}10] direction. }
    \label{fig:structure}
\end{figure}

\textit{Ab initio} calculations were performed using density-functional theory as implemented
in the Vienna \textit{ab initio} simulation package ({\sc vasp}).~\cite{kresse1993,kresse1996}
Spin-orbit coupling was included in the band structure calculations. The local-density approximation
(LDA) was employed for the exchange-correlation functionals to optimize the lattice structures.
In the band structure calculations and MLWF projections, we went beyond the LDA and adopted the
hybrid functional method.~\cite{paier2006,heyd2004,heyd2005} We studied a HgTe (001)
surface that was grown in recent experiments.~\cite{bruene2011,yao2013} Our optimized
equilibrium lattice constant was 6.438 \AA, which is close to the experimental value of 6.461 \AA
(Ref.~\onlinecite{roessler2008}). On the basis of the optimized cubic lattice [Fig.~\ref{fig:structure}(a)],
we adopted the primitive tetragonal unit cell [Fig.~\ref{fig:structure}(b)]
to simulate the (001) surface. The substrate-induced in-plane strains are
applied as $\epsilon_x = \epsilon_y = - \epsilon_z = \epsilon$ for this isotropic cubic system,
where $0 \leq \epsilon \leq 3\%$. The CdTe-substrate-induced strain is equivalent to $\epsilon = 0.3\%$
according to a recent experiment.~\cite{bruene2011} For each strain value, we extracted the corresponding
MLWFs from the bulk calculations and then calculated the surface states, where the Bloch wave functions
were projected to the Hg-$s$ and Te-$p$ orbitals.

\begin{figure}[htb]
    \includegraphics[width=0.9\linewidth]{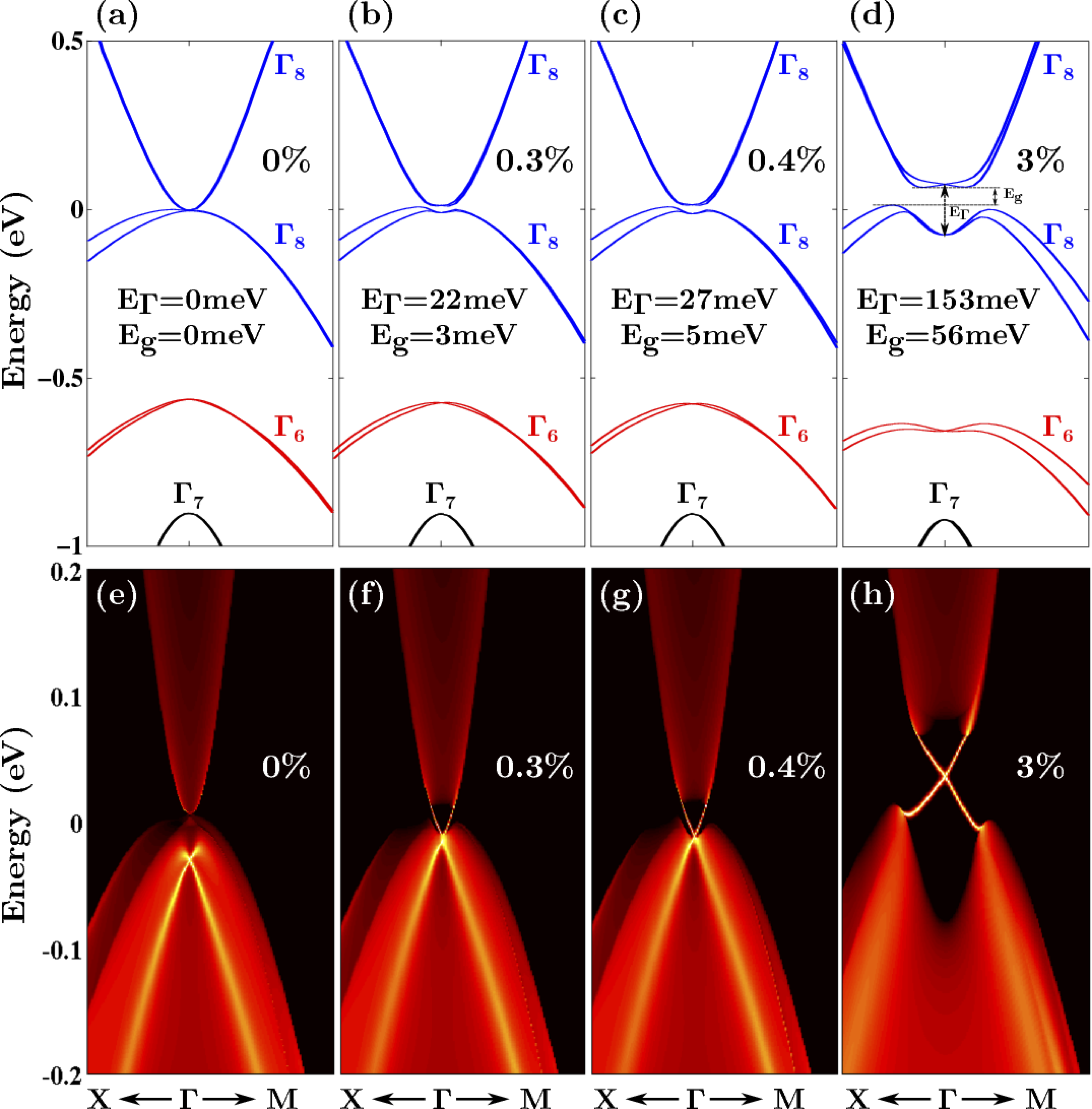}
    \caption{Bulk band structures of HgTe at (a) 0\%, (b) 0.3\% (experimental value), (c) 0.4\%, (d) 3\% strain. $E_{\rm g}$ is the indirect gap, and $E_{\Gamma}$ is the direct gap at the $\Gamma$ point.
    Surface LDOS of HgTe (001) surface at (e) 0\%, (f) 0.3\% (experimental value), (g) 0.4\%, (h) 3\% strain. Bright (white) regions stand for surface states; dark (red) regions represent bulk states. The 2D surface Brillouin zone corresponds to the tetragonal unit cell shown in Fig. 1(b), where $X$ represents the [110] direction, and $M$ represents the [010] direction with respect to the cubic lattice. }
    \label{fig:GF}
\end{figure}

\begin{figure*}[htb]
    \includegraphics[width=0.9\linewidth]{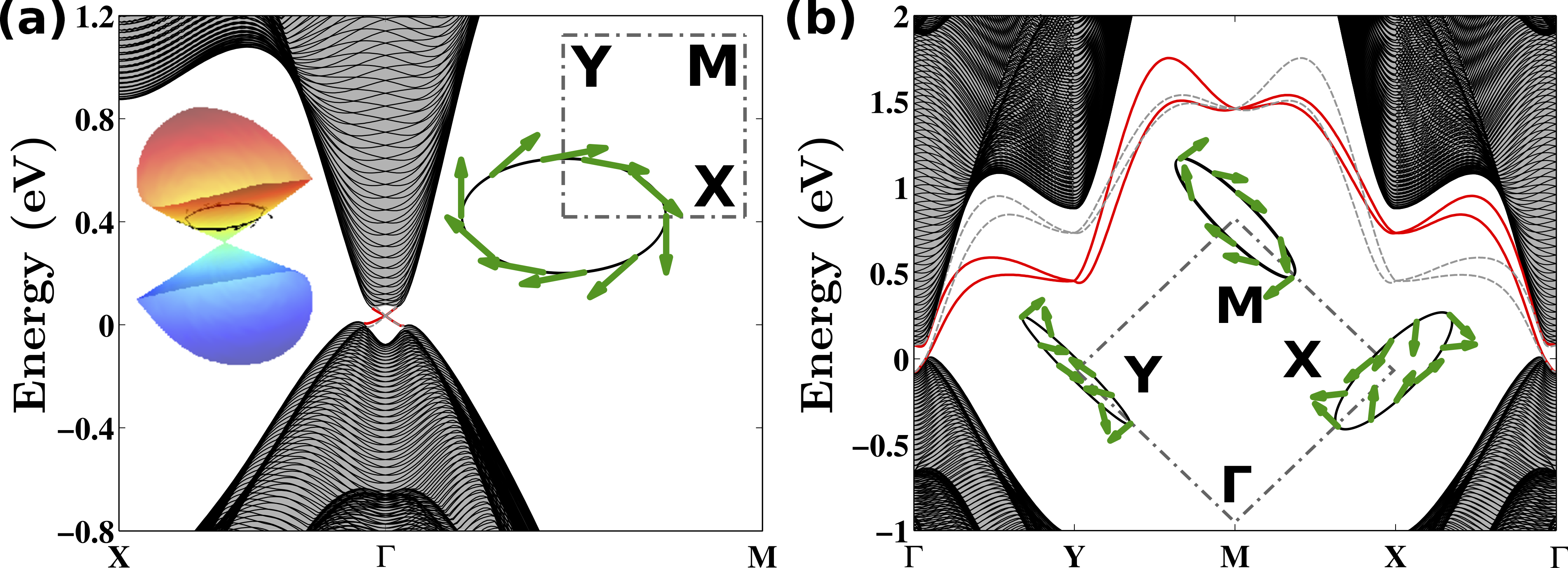}
    \caption{Band structure for 3\% strained HgTe of finite surfaces (a) terminated by Te atoms and (b) terminated by Hg atoms for 240 atomic layers (about 38~nm in thickness).
             Solid red line is contributed from top layer; dashed gray line is contributed from bottom layer.
             Spin direction is labeled for the topological states of the Fermi surface of the top layer, which is the upper Fermi energy.
             Gray shading represents bulk states.}
    \label{fig:Te}
\end{figure*}


\section{results and discussion}

The bulk band structure of HgTe is illustrated in Fig.~\ref{fig:GF}(a)--\ref{fig:GF}(d). 
The strain-free lattice is a zero-gap semimetal, in which the conduction and valence bands
are degenerate at the $\Gamma$ point. They are called $\Gamma_8$ bands according to
their symmetry and are contributed by Te-$5p$ states. Another valence Hg-$6s$ band,
called $\Gamma_6$, lies below $\Gamma_8$; this is treated as an ``inverted'' band
structure with respect to that of a normal insulator such as CdTe.~\cite{bernevig2006d} When
uniaxial strain breaks the cubic symmetry, the $\Gamma_8$ degeneracy is removed,
inducing an energy gap. Both unstrained and strained lattices preserve the band
inversion and thus host TSSs. Figures~\ref{fig:GF}(a)--\ref{fig:GF}(d) show the evolution
of the bulk band structures around the $\Gamma$ point with respect to the external strains.
When strain exists, the degenerate $\Gamma_8$ bands split and create a direct energy
gap at $\Gamma$ ($E_{\Gamma}$) and an indirect energy gap near the $\Gamma$ point
($E_{\rm g}$). Both $E_{\Gamma}$ and $E_{\rm g}$ increase monotonously with increasing
strain. Further, the dispersion of the $\Gamma_6$ bands is also slightly modified by the
strain because of symmetry breaking. Note also that hybrid functional
calculations reproduced the correct band ordering, $\Gamma_7 < \Gamma_6 <\Gamma_8$, whereas
the LDA tends to overestimate the band inversion and yields $\Gamma_6 < \Gamma_7 <\Gamma_8$.
Hence, we believe that hybrid functionals give more accurate band structures than the LDA
does for HgTe, consistent with recent calculations.~\cite{svane2011,cardona2009}

To demonstrate the surface states, we calculated the local density of states
(LDOS) using the Green's function method: LDOS $(k,E)$=$-\frac{1}{\pi}TrImG_{00}(k,E)$. 
On the semi-infinite surface, $G_{00}$ is the Green's function projected from the 3D
bulk to the top layer~\cite{sancho1984}, which is constructed from the tight-binding
Hamiltonian based on MLWFs.
The surface band structures are shown in Figs.~\ref{fig:GF}(e)--\ref{fig:GF}(h). Without
strain, there are already two types of TSSs. One is at the conduction band edge and
is slightly decoupled from the bulk states, and the other is below the top valence band
with Dirac-type crossing and is mixed with the bulk states in energy [Fig.~\ref{fig:GF}(e)].
This feature agrees well with previous calculations using a model based on a six-band
Kane Hamiltonian.~\cite{chu2011} When the strain opens an energy gap, the higher surface
states still remain inside the gap, whereas the lower Dirac cone moves up [Fig.~\ref{fig:GF}(f)].
At a critical strain of 0.4\% [Fig.~\ref{fig:GF}(g)], two types of TSSs merge together 
with the Dirac point at the top of the valence band. Subsequently, a full Dirac cone forms
inside the gap for even larger strains [Fig.~\ref{fig:GF}(h)]. In the experiment, the HgTe
thin film bears a 0.3\% strain from the CdTe substrate. Thus, its band structure is
expected to be similar to our 0.3\% uniaxial strain case [Fig.~\ref{fig:GF}(f)], where
the Dirac point exists inside the bulk conduction bands. Even the lower Dirac surface
states overlap with the bulk bands in energy; their linear dispersion extends clearly
from the $\Gamma$ point to higher momentum in the 2D Brillouin zone, which indicates
that these TSSs are well separated from the bulk states in real space. This is also
consistent with recent ARPES measurements.~\cite{bruene2011,yao2013,crauste2013} We note
that it might be necessary to take these two types of surface states into account when
interpreting future ARPES and transport experiments.

In calculations of the above-surface LDOS, we adopted Te sites as the top layer of the
semi-infinite surface. It is also possible to have a Hg layer as the outer surface.
However, the Te-terminated model provides a surface band structure that is much closer
to that obtained experimentally using ARPES than that of the Hg-terminated one, as we will see. Note
that this does not necessarily mean that the real experimental surface is terminated
by Te. Actually, our fully \textit{ab initio} calculations based on a slab model with
either Te or Hg terminations yield surface states that differ from those obtained experimentally,
where several trivial dangling bond states appear together with TSSs (also see Ref.~\onlinecite{virot2013}).
However, these dangling bond states were not reported in
experimental results,~\cite{bruene2011,yao2013,crauste2013} which may indicate a cleaved
surface in ARPES with various coexisting terminations. This is similar to the
TlBiSe$_2$ type of TIs, where tight-binding calculations~\cite{yan2010} predicted
surface states more consistent with experiments~\cite{sato2010,kuroda2010,chen2010}
than those from \textit{ab initio} simulations,~\cite{lin2010b} in which mixed Tl
and Se terminations were revealed recently by scanning tunneling microscopy.~\cite{Kuroda2013}

The Te- and Hg-terminated surfaces can act as typical examples to illustrate the
robustness and flexibility of TSSs. We constructed two slab models 240 atomic layers
thick ($\sim$38 nm) with top and bottom surfaces. One slab is terminated by Te on both
surfaces, and the other is terminated by Hg. To exhibit a simple Dirac cone with a large
bulk energy gap, we adopted the MLWFs from 3\% strained bulk HgTe. The slab Hamiltonians
were directly diagonalized to obtain the band structures (see Fig.~\ref{fig:Te}). Take the
Te-terminated surface as an example. Although it behaves approximately isotropically at smaller
strains [Figs.~\ref{fig:GF}(a)--\ref{fig:GF}(c)], the surface state of the Dirac cone exhibits considerable
anisotropy when it shifts inside the bulk gap at a large strain [Fig.~\ref{fig:Te}(a)].
TSSs are more dispersive along the $\Gamma-Y$ direction on the top surface because surface
Te atoms are interconnected by the second layer of Hg atoms to form a nearly 1D chain along the
$y$ direction showing in Fig.~\ref{fig:structure}(c). Because the top and bottom surfaces are not equivalent owing to the lack of
inversion symmetry, the surface states on both surfaces are non-degenerate in the band structure.
We highlighted the TSSs on the top surface by solid red lines and those on the bottom surface by
dashed gray lines in Fig.~\ref{fig:Te}. In Fig.~\ref{fig:Te}(b), one can see that the $\Gamma-Y$ dispersion of the top
TSSs is equivalent to the $\Gamma-X$ dispersion of the bottom TSSs, because the 1D Te chains
from the top and bottom surfaces are perpendicular to each other. Further, the Te-terminated
surface exhibits a single Dirac cone at the $\Gamma$ point as in Fig.~\ref{fig:GF}(h),
whereas the Hg-terminated surface shows three Dirac cones at other time-reversal-invariant
k-points, $Y$, $M$, and $X$. This indicates that the energy dispersion of the TSSs responds quite
flexibly to local surface potential modifications, whereas the existence of TSSs is always
protected by the underlying topology of the bulk band structure. We note that the surface
states of the Hg-terminated surface are similar to previous calculations in Refs.~\onlinecite{dai2008,yan2012}.
For the Te surface, one can observe a left-handed spin texture in the upper Dirac cone. For the
Hg surface, there exist two right-handed ($X$ and $Y$) vortices and one left-handed ($M$) vortex.
Because an odd number of Dirac cones exist on either the Te or Hg surface, an electron always
obtains a nontrivial Berry phase, $\pi$ or $-\pi$, due to the spin vortices when it circles
around the Fermi surface, showing a topological feature. In addition, one can
find that the Te-terminated surface exhibits the same left-handed spin texture as
Bi$_2$Se$_3$.~\cite{liu2010,hsieh2009}

\section{conclusions}
In summary, we studied the TSSs on the (001) surface of strained HgTe.
The position and number of Dirac points is sensitive to whether the surface terminates in Te or Hg atomic layers.
The Te-terminated surface exhibits a simple Dirac cone at the $\Gamma$ point, whereas the Hg-terminated
surface has three Dirac points, at the $X$, $Y$ and $M$ points. On the basis of our calculations for different
strain values, we expect that the Dirac point lies slightly below the bulk valence band maximum
for the substrate-strained HgTe in previous experiments.
Regardless of the surface termination, the spin textures of the Dirac cones carry a Berry phase $\pi$,
which indicates a topological feature.


\begin{acknowledgments}
We are grateful to C.-X. Liu for fruitful discussion. B.Y. and C.F. acknowledge financial support in the form of an ERC Advanced Grant (291472).
\end{acknowledgments}

\end{document}